\documentclass[article,twocolumn,american,amsmath,amssymb,aps,physrev,prb,prapplied,floatfix,longbibliography,superscriptaddress]{revtex4-2}
\usepackage{mathptmx}

\usepackage{color} 

\usepackage{fontenc}
\usepackage{graphicx}
\usepackage{dcolumn}
\usepackage{bm}
\usepackage{siunitx}
\usepackage[letterpaper]{geometry}
\usepackage{esint}
\PassOptionsToPackage{normalem}{ulem}
\usepackage{ulem}
\usepackage{natbib}
\usepackage{psfrag}
\usepackage{hyperref}
\usepackage{diagbox}
\usepackage{multirow}
\usepackage{longtable}
\usepackage{booktabs}
\usepackage[caption=false]{subfig}
\definecolor{darkpastelgreen}{rgb}{0.01, 0.75, 0.24}
\makeatletter

\begin{document}

\title{Energy Barriers for Thermally Activated Magnetization Reversal in Perpendicularly Magnetized Nanodisks in a Transverse Field}

\begin{abstract}
Thermally-induced transitions between bistable magnetic states of magnetic tunnel junctions (MTJ) are of interest for generating random bitstreams and for applications in stochastic computing. An applied field transverse to the easy axis of a perpendicularly magnetized MTJ (pMTJ) can lower the energy barrier ($E_b$) to these transitions leading to faster fluctuations. In this study, we present analytical and numerical calculations of $E_b$ considering both coherent (macrospin) reversal and non-uniform wall-mediated magnetization reversal for a selection of nanodisk diameters and applied fields. Non-uniform reversal processes dominate for larger diameters, and our numerical calculations of $E_b$ using the String method show that the transition state has a sigmoidal magnetization profile. The latter can be described with an analytical expression that depends on only one spatial dimension, parallel to the applied field, which is also the preferred direction of profile motion during reversal. Our results provide nanodisk energy barriers as a function of the transverse field, nanodisk diameter, and material characteristics, which are useful for designing stochastic bitstreams.

\end{abstract}

\author{Corrado C. M. Capriata}
\email{capriata@kth.se}
\affiliation{Division of Electronics and Embedded Systems, KTH - Royal Institute of Technology, Stockholm, Sweden}%
\affiliation{Center for Quantum Phenomena, Department of Physics, New York University, New York, NY 10003, USA}

\author{B. Gunnar Malm}
\affiliation{Division of Electronics and Embedded Systems, KTH - Royal Institute of Technology, Stockholm, Sweden}%

\author{Andrew D. Kent}
\affiliation{Center for Quantum Phenomena, Department of Physics, New York University, New York, NY 10003, USA}

\author{Gabriel D. Chaves-O'Flynn}
\email{gchaves@ifmpan.poznan.pl}
\affiliation{Institute of Molecular Physics, Polish Academy of Sciences, Poznań, Poland}%


\date{\today}

\maketitle

\section{\label{sec:intro}Introduction}
There has been a recent surge in interest in innovative computational approaches that mimic the flexibility of neural systems \cite{darwish_survey_2020, shastri_photonics_2021, hoffmannQuantumMaterialsEnergyefficient2022}. Proposals based on magnetic systems include neuromorphic computing \cite{grollier_neuromorphic_2020}, reservoir  \cite{tanaka_recent_2019}, and stochastic computing \cite{liu_survey_2021}. For these applications, it is necessary to find physical systems with true random behavior \cite{qu_variation-resilient_2018,bassham_statistical_2010, rehm_stochastic_2023}. Magnetic tunnel junction devices can exhibit random two-state fluctuations. However, magnetic tunnel junctions have primarily been used for traditional information storage, where the state of the information bit must remain stable over time and be resistant to thermal fluctuations for decades~\cite{augustine2010, pajouhi2015}. To achieve this goal, perpendicular magnetic tunnel junctions (pMTJs) have been extensively studied and optimized as memory elements~\cite{puebla_spintronic_2020, sun_metrology_2022}. In contrast, devices suitable for random number generation require energy barriers that can be easily overcome by thermal perturbations. Fortunately, pMTJs can also be designed to generate random numbers with a high rate of switching between two equally likely metastable states \cite{vodenicarevic_low-energy_2017, parks_superparamagnetic_2018, kaiser_subnanosecond_2019, hayakawa_nanosecond_2021, safranski_demonstration_2021,rehm_stochastic_2023}.

Kramers' theory for thermally activated escape over a potential barrier \cite{hanggi_reaction-rate_1990, Braun_1994} is useful in this regard. This theory states that the rate of thermally induced switching between two metastable states, $\Gamma$, obeys an Arrhenius law $\Gamma=\Gamma_0 \exp(-E_b/(k_BT))$, where $\Gamma_0$ is an attempt frequency, typically in the GHz range, $k_B$ is Boltmann's constant and $T$ is the temperature. $E_b$ is the energy barrier, the difference between the energy of the transition state and the energy of the metastable state, where the transition state corresponds to the lowest energy saddle point between the two metastable states.

Previous research on randomly fluctuating magnetic tunneling junctions (MTJs) has primarily focused on easy-plane systems, which have been considered the faster alternative~\cite{bordersIntegerFactorizationUsing2019, camsari_double-free-layer_2021}. However, in this article, we present a theoretical model for MTJs with uniaxial magnetic anisotropy in the presence of a transverse field. In fact, in pMTJs, applying a transverse magnetic field lowers the energy barrier, which in turn increases the fluctuation rates~\cite{Chen_2022}.

We investigated perpendicularly magnetized nanodisks, that correspond to the so-called free layer of a perpendicular magnetic tunnel junction (pMTJ) memory element. Our starting point is a macrospin model developed by Garanin in Ref.~\cite{garaninThermallyActivatedEscape1999}. However, our analysis accounts for non-uniform magnetization switching, which is characteristic for nanodisks larger than a certain minimum size. The theory presented in this study predicts how $E_b$ varies with nanodisk diameter and the magnitude of the applied field. To validate our analytical predictions, we compared them against numerical results obtained using the String method~\cite{eStringMethodStudy2002, eSimplifiedImprovedString2007} implemented in OOMMF micromagnetic simulator~\cite{donahue_oommf_1999}. This technique has been previously used by our group to study thin films~\cite{chaves-oflynn_thermal_2015}, nanomagnets~\cite{chaves-oflynn_energy_2013}, and nanorings~\cite{martensMagneticReversalNanoscopic2006, chaves-oflynn_micromagnetic_2009}.

The paper is organized as follows. In Section \ref{sec:setup}, we describe the structure we model and simulate. In Section \ref{sec:macrospinmodel}, we present the macrospin model and the micromagnetic simulation results used to test it. Section \ref{sec:micromagmodel} covers the non-uniform reversal model and the String method technique used to assess the model. The results of the String method simulation are presented and discussed in detail in Section \ref{sec:results}. The paper ends with a summary in Section \ref{sec:summary}.

\section{\label{sec:setup}Free layer geometry}
\begin{figure}
\includegraphics[width=\columnwidth]{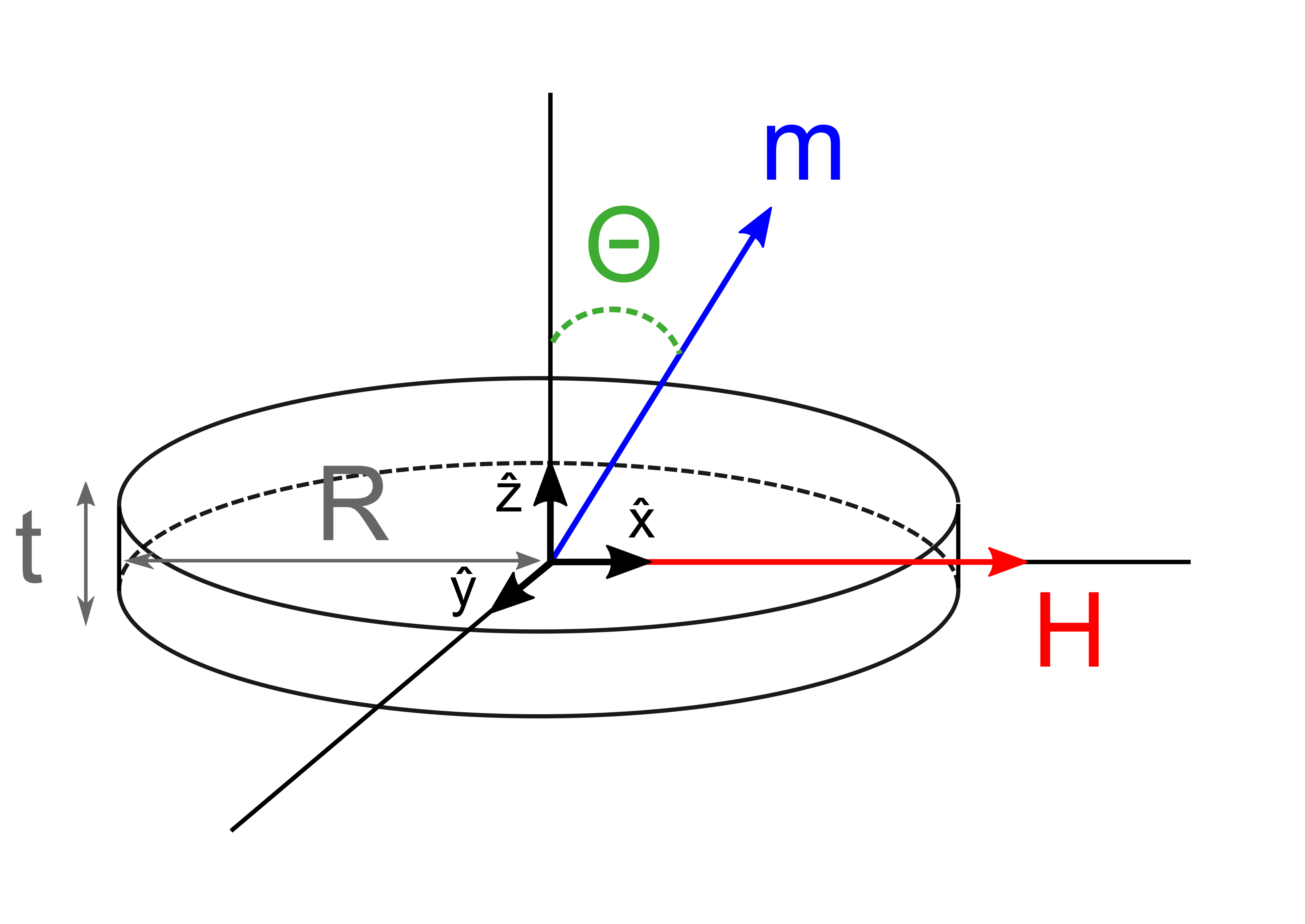}
    \caption{Schematic of the ferromagnetic disk under consideration. It has radius $R$, thickness $t$ and an easy magnetic axis in the z-direction (i.e., perpendicular magnetic anisotropy). The field is applied in-plane along the x-direction.
    } 
    \label{fig:demagtensor}
\end{figure}

A schematic of a pMTJ free layer is shown in Fig. \ref{fig:demagtensor}. It is modeled by a ferromagnetic nanodisk with geometrical and magnetic properties chosen to have net magnetic anisotropy perpendicular to the film plane. The coordinate system has $\bm{\hat{x}}$ parallel to the applied magnetic field and $\bm{\hat{z}}$ perpendicular to the film plane. The magnetocrystalline anisotropy axis $\mathbf{\hat{k}}$ is parallel to $\mathbf{\hat{z}}$. To describe the magnetization, we use spherical coordinates in which $\Theta$ is the angle of the magnetization from the $\bm{\hat{z}}$ axis. Because of the orientation of the field,  the magnetization lies in the $x-z$ plane.

We use SI units and dimensionless expressions are used to simplify the algebra and provide generalizations. Dimensionless quantities are expressed in lower case if there is no ambiguity ($\mathbf{m}=\mathbf{M}/M_s$), or with tildes if necessary (as in $\tilde{x}=x/l_\mathrm{ex}$).

\section{\label{sec:macrospinmodel}Macrospin model}
For magnetization reversal in very small samples coherent rotation of the magnetization (macrospin reversal) can be favored, while in samples larger than the exchange length there may be spatial variations of the magnetization (non-uniform reversal). Thus, we start here with the macrospin model applied to uniformly magnetized thin ferromagnetic disks with radius $R$, thickness $t$, and volume $V$.

The magnetic energy density, $\mathcal{E}=E/V$, is obtained from the sum of dipolar ($\mathcal{E}_\mathrm{d}$), anisotropy ($\mathcal{E}_\mathrm{k}$), and Zeeman ($\mathcal{E}_\mathrm{Z}$) energy densities:
\begin{align}
    \mathcal{E}(\mathbf{m})& 
     =\mathcal{E}_\mathrm{d}
     +\mathcal{E}_\mathrm{k}
     +\mathcal{E}_\mathrm{Z}.
\end{align}
The individual terms depend on $\mathbf{m}$ as follows:
\begin{align}
    \mathcal{E}_\mathrm{d}& 
    =\zeta K_d (\mathbf{m\cdot\hat{z}})^{2}
    \\
    \mathcal{E}_\mathrm{k} & =-K(\mathbf{m\cdot\hat{k}})^{2}\\
    \mathcal{E}_\mathrm{Z} & =-\mu_{0}M_{s}\mathbf{H}\cdot\mathbf{m},
\end{align}
where $M_s$ is the saturation magnetization, $K$ is the magnetic anisotropy, $K_d=\mu_{0}M_{s}^{2}/2$ is the demagnetization ﬁeld energy coeﬃcient, and $\mathbf{H}=H_{x}\mathbf{\hat{x}}$ is the applied in-plane
magnetic field. 

The coefficient $\zeta$ is obtained from the disk magnetostatic tensor using $\zeta = (3N_{zz}-1)/2$. Both are purely geometric factors that depend only on the disk aspect ratio and are close to unity if $(t/R) \ll 1$ (see Refs. \cite{beleggiaEquivalentEllipsoidMagnetized2006,chaves-oflynn_thermal_2015}
and Appendix~\ref{app:demagnetizingfactor}).
The applied field will be normalized using the effective perpendicular anisotropy of the layer $\mathbf{h}\equiv\mathbf{H}/[M_{s}(Q-\zeta)]$ with $Q=K/K_d$. 

After expressing the magnetization unit vector in polar coordinates $\mathbf{m}=\left(\cos\Phi\sin\Theta,\sin\Phi\sin\Theta,\cos\Theta\right)$ and energy densities in terms of $\mathcal{E}_0=K_d(Q-\zeta)$ we obtain a rescaled energy density, $\tilde{\mathcal{E}}=\mathcal{E}/\mathcal{E}_0$:
\begin{equation}\label{eq:energydensity}
    \tilde{\mathcal{E}}=-\cos^{2}\Theta-2h\sin\Theta\cos\Phi.
\end{equation}

Because the calculation of energy barriers is the main goal of this work, we emphasize our conventions at this point. The total energy is obtained after integrating over the volume of the disk and will be denoted by the unscripted letter $E$. The problem characteristic energy, $E_0$, will be introduced in Section \ref{sec:micromagmodel}. Energy barriers, $E_b$, will be expressed in units of $k_BT$ (with $T=\SI{300}{\kelvin}$). Energy density terms are denoted with script character $\mathcal{E}$ and are of the order of $1\ \mathrm{MJ/m^3}$. Dimensionless energy densities are denoted with tilde $\tilde{\mathcal{E}}$ after normalization by $\mathcal{E}_0$. A summary of the conversion relations is provided in Table~\ref{tab:Conversion-relations}. We now proceed to find the extremals of the energy.

\begin{table*}
\caption{\label{tab:Conversion-relations}Summary of conversion relations for dimensionless units.}
\begin{longtable}[c]{c>{\centering}p{1.5in}>{\centering}p{2in}c}
\toprule 
Symbol & Variable & Conversion Relation & Normalization Quantity\tabularnewline
\midrule
\midrule 
$E$ & Energy & $E=E_{0}\tilde{E}$ & $E_{0}=At$\tabularnewline
\midrule 
$\mathcal{E},\mathcal{E}_{d},\mathcal{E}_{k},\mathcal{E}_{Z},\mathcal{E}_{\mathrm{ex}}$ & Energy density terms, in order: total, dipolar, anisotropy, Zeeman,
exchange. & $\mathcal{E}=\text{\ensuremath{\mathcal{E}_{0}}\ensuremath{\tilde{\mathcal{E}}}}$ & $\ensuremath{\mathcal{E}_{0}}=K_{d}(Q-\zeta)$\tabularnewline
\midrule 
$\mathbf{H}=H\mathbf{x}$ & Applied external field. & $H=hM_{s}(Q-\zeta)$ & $M_{s}(Q-\zeta)$\tabularnewline
\midrule 
$\tilde{x},\tilde{y}$ & In plane dimensionless coordinates. & $x=l_{\mathrm{ex}}\tilde{x}$ & $l_{ex}=\sqrt{\frac{2A}{\mu_{0}M_{s}^{2}\left(Q-\zeta\right)}}$\tabularnewline
\bottomrule

\end{longtable}

\end{table*}

The transition and metastable states of the energy are magnetization directions where the effective field, $\frac{\delta\mathcal{E}}{\delta \mathbf{m}}$, vanishes: 
\begin{align}
\frac{\partial\tilde{\mathcal{E}}}{\partial \Theta}=
-\cos\Theta \left(\sin\Theta +h\cos\Phi\right) 
=0\label{eq:thetafield}
\\
\frac{1}{\sin \Theta}
\frac{\partial\tilde{\mathcal{E}}}{\partial \Phi}=-h\sin\Phi=0
\label{eq:phifield}
\end{align}
From Eq.~\ref{eq:phifield} we conclude that $\Phi=0$. There are four uniform solutions of Eq.~\ref{eq:thetafield}. The first two
\begin{equation}
    \Theta_{\pm}=\pm\frac{\pi}{2};\qquad \tilde{\mathcal{E}}_{\pm}=\mp 2h,
\end{equation}
are the polar angles of the energy maximum and saddle. There are also two mutually supplementary angles that share the same energy
\begin{equation}
\label{eq:theta}
    \Theta_{\uparrow,\downarrow}=\sin^{-1}h\qquad \tilde{\mathcal{E}}_{1}=-\left(1+h^2\right),
\end{equation}
and correspond to energy minima. In Garanin's macrospin reversal model \cite{garaninThermallyActivatedEscape1999}, the energy barrier for the most likely thermally activated transition is given by:

\begin{equation}
    \tilde{\mathcal{E}_b}=\tilde{\mathcal{E}}_+-\tilde{\mathcal{E}}_1=\left(1-h\right)^2=\frac{E_b}{K_dV(Q-\zeta)}
\end{equation}
For the reader's convenience, we write $E_b$ explicitly in dimension-full units.

\begin{equation}
    \label{eq:shapefactor}
    E_b=\left(1-
    \frac{H}{M_{s}(Q-\zeta)}
    \right)^2K_dV(Q-\zeta)
\end{equation}

Our model slightly refines Garanin's prediction by explicitly considering shape anisotropy effects caused by the magnetostatic interaction which are captured here in the size-dependent factor $(Q-\zeta)$. 

\subsection{Micromagnetic Calculations}
To exemplify the predictions of the macrospin we perform pairs of overdamped micromagnetic simulations (damping factor $\alpha=1$) using $\mathbf{m}\approx\mathbf{\hat{x}}$ as the initial configuration. For these simulations, the initial magnetization was set marginally out of the $x-y$ plane and in opposite directions so that the magnetization relaxes to different metastable states. The parameters for these simulations were set to match those of our previous experimental studies \cite{Rehm2019}. They are as follows: Heisenberg exchange constant $A=$\SI{4.2}{\pico\joule\per\meter}, saturation magnetization $M_s=$\SI{0.58}{\mega\ampere\per\meter}, bulk magnetic anisotropy $K=$\SI{0.39}{\mega\joule\per\meter^3}. The material constant becomes $Q=1.84$. The cells in the simulation had dimensions: $2.5\times2.5\times2.6~\mathrm{nm^3}$.

\begin{figure}
    \includegraphics[width=\columnwidth]{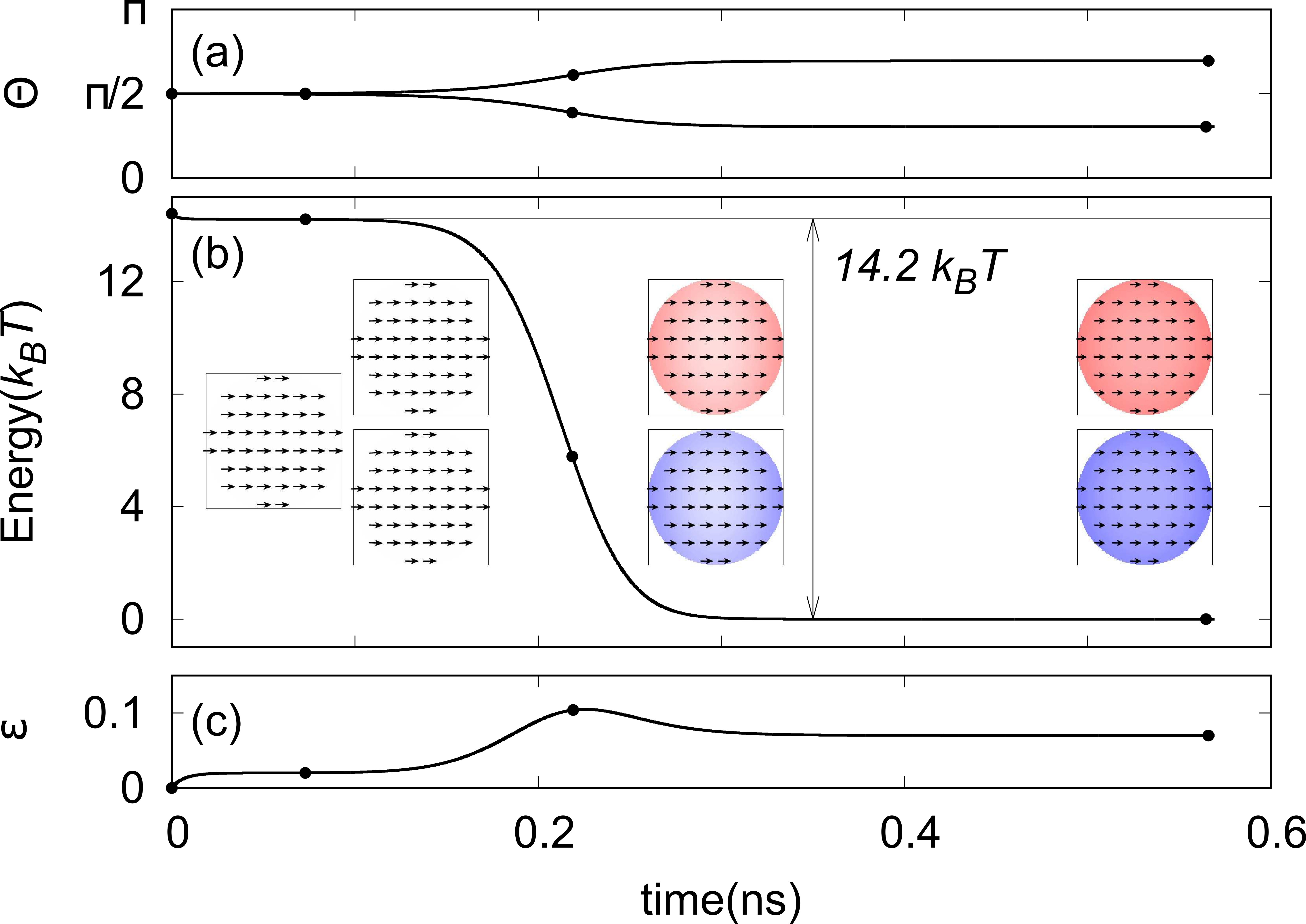}
    \caption{\label{fig:overdamped} 
    (a) Two overdamped micromagnetic simulations of a \SI{60}{\nano\meter} device at a field of \SI{0.6}{\tesla} that relax to different minima. The energy barrier is measured directly from the difference between the initial and final states (b). The slow initial evolution confirms that the macrospin saddle is a critical point of the energy landscape. The two minima are symmetric and follow identical curves. (c) The non-uniformity as defined in the text remain low and overlapping for the two simulations.
    }
\end{figure}

A typical result from the overdamped simulations is shown in Fig.~\ref{fig:overdamped}. The difference between the two traces can only be seen in the plot of $\Theta(t)$, Fig.~\ref{fig:overdamped}(a), because the energy and non-uniformity results are identical. This behavior can be taken as confirmation that the initial magnetization is indeed a saddle point of the energy landscape. The initial magnetization dynamics are at first very slow before decaying to either of the energy minima, as is evident in the plateau of the energy vs time curve of Fig.~\ref{fig:overdamped}(b).
 
We use the spatial average of each magnetization component to quantify the non-uniformity of the configuration, $\epsilon=\sqrt{1-\bar{m_x}^2-\bar{m_y}^2-\bar{m_z}^2}$. As can be seen in Fig.~\ref{fig:overdamped}(c), this quantity remains small during the overdamped relaxation. This fact supports our use of overdamped dynamics to validate macrospin predictions.

\begin{figure}
    \includegraphics[width=\columnwidth]{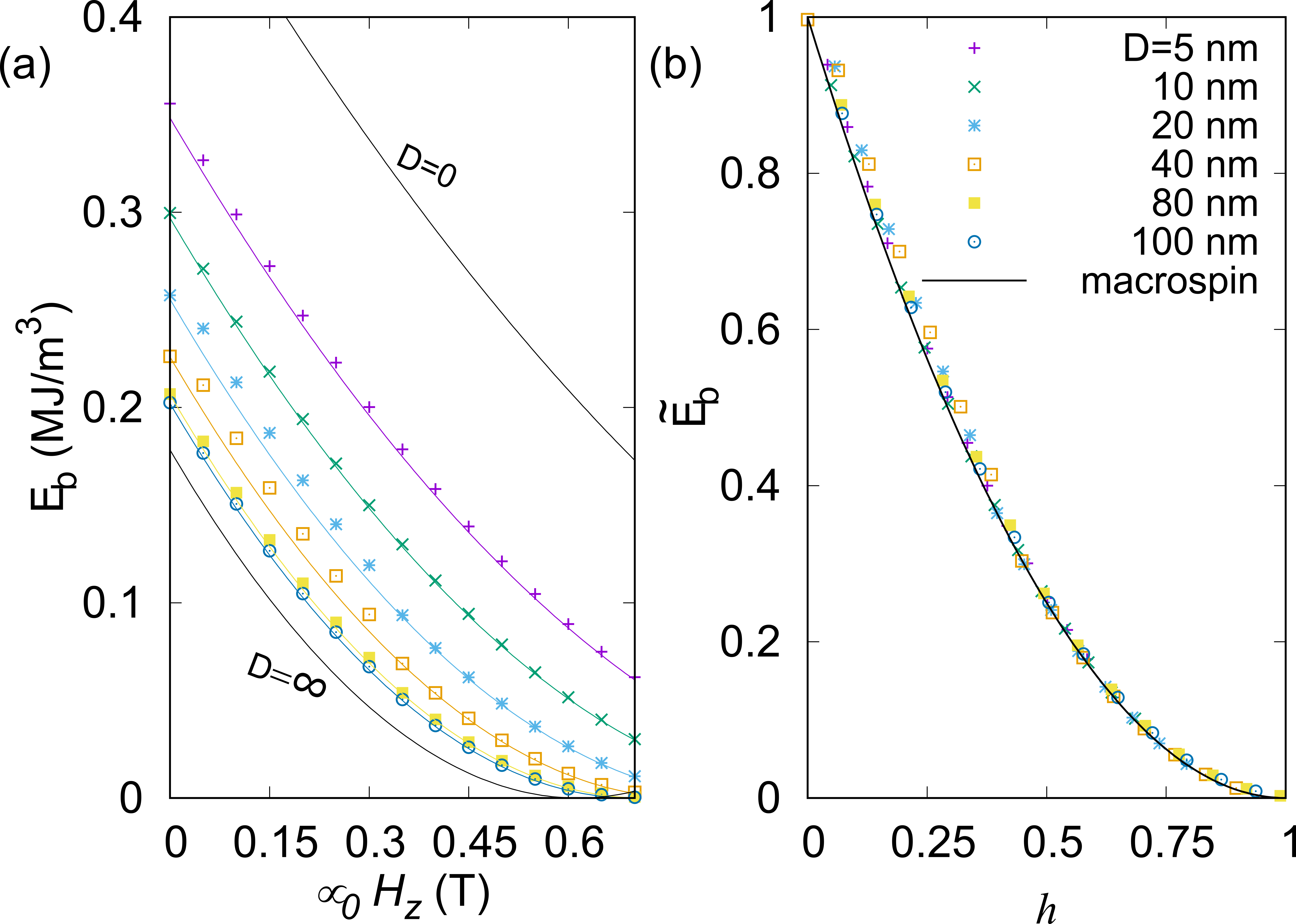}
    \caption{\label{fig:overdampedResults} (a) Energy barrier $\mathcal{E}_b$ vs field $\mu_0H_z$ without the shape factor correction $(Q-\zeta)$. The black curves show the limiting values for $D=0$ and $D\rightarrow\infty$. Points are from micromagnetic simulation, continuous lines are obtained from Eq.~\ref{eq:shapefactor}. (b) Non-dimensional energy barriers, $\tilde{\mathcal{E}_b}$, vs scaled field, $h$. All results from overdamped simulations collapse onto the macrospin model after appropriate normalization.
    }  
\end{figure}

From a systematic repetition of this procedure, we determine the energy barrier versus applied transverse field for six distinct diameters in the range $5 < D < 100$ \SI{}{\nano\meter}. The energy barrier and field are scaled as indicated in Eq.~\ref{eq:shapefactor}. In consequence, all data appear on a single curve Fig.~\ref{fig:overdampedResults}(b). This result confirms the universality of Garanin's macrospin approximation, provided the proper shape-induced magnetization corrections are taken into account. 

For comparison, Fig.~\ref{fig:overdampedResults}(a) shows the same data in SI units. An immediate consequence of these studies is the counter-intuitive result observable here. It is generally expected that the energy barrier of a uniformly magnetized system increases with volume. However, it is shown that the energy barrier for fixed disk thickness actually decreases with the radius for any fixed field. This is because the effective perpendicular anisotropy weakens as $R/t\rightarrow \infty$. In Fig.~\ref{fig:overdampedResults}(a), the simulation data points do not exactly follow the theoretical lines. We did not find a clear relation between the offset between the macrospin model and the theory and the radius of the device. Therefore we attribute the mismatch to effects of the numerical discretization.

The decrease in the energy barrier occurs even under the assumption of uniform magnetization and the curves seem to reach convergence at $D\approx$ \SI{100}{\nano\meter}. Nonetheless, this limiting energy barrier is much larger than a barrier obtained from domain wall mediated reversal, as will be introduced in the next section.

\section{\label{sec:micromagmodel}Domain Wall Mediated Reversal}

While the macrospin model is a useful approximation for device sizes smaller than the exchange length, it is natural to expect that it fails once the system is large enough to accommodate variations of the magnetization. To consider this scenario, we rewrite the energy as a surface integral with four spatially varying contributions:
\begin{align}\label{eq:nonuniformmicromagneticenergy}
    E =t\int\mathcal{E}(\mathbf{m})d^{2}\mathbf{r}
     =t\int\left(
     \mathcal{E}_\mathrm{ex}
     +\mathcal{E}_\mathrm{d}
     +\mathcal{E}_\mathrm{k}
     +\mathcal{E}_\mathrm{Z}
     \right)d^{2}\mathbf{r}.
\end{align}
The energy now includes the exchange energy 
\begin{align}
    \mathcal{E}_\mathrm{ex} & 
    =A\left|\nabla\mathbf{m}\right|^2=A\left(
    \left|\nabla\Theta\right|^{2}
    +\sin^2\Theta\left|\nabla\Phi\right|^{2}
    \right)
\end{align}
where $A$ is the Heisenberg exchange constant.

 Equation \ref{eq:nonuniformmicromagneticenergy} estimates the dipolar interaction using the magnetostatic tensor of uniformly magnetized disks. It equals the dipolar interaction between a uniformly magnetized disk and a small differential volume, $td^2r$, with the same magnetization. As such, only the surface charges are accounted for, at the top and bottom surfaces of the disk, and at the in-plane edges. This rough approximation neglects the role of bulk magnetic charges but is a necessary step towards a local energy functional amenable to analysis with variational calculus. A rigorous justification is beyond the scope of this paper, but the leading order corrections to this approach are usually very small (see, e.g., Ref. \cite{kohn_another_2005,gioia_micromagnetics_1997,muratov_domain_2017,martensMagneticReversalNanoscopic2006}).

To obtain non-dimensional expressions we rescale the in-plane spatial coordinates ($\mathbf{r}=l_\mathrm{ex}\mathbf{\tilde r}$) by an exchange length, 
\begin{equation}\label{eq:exchangelength}
    l_{\mathrm{ex}}=\sqrt{\frac{2A}{\mu_{0}M_{s}^{2}(Q-\zeta)}};
\end{equation}
and the energies with a characteristic value, $E_0=At$;

The energy density rescaling factor can now be written as $\mathcal{E}_0=E_0/(l_\mathrm{ex}^2t)$ and the 
dimensionless energy density  ($\tilde{\mathcal{E}}=\mathcal{E}/\mathcal{E}_0$)
becomes:
\begin{equation}\label{eq:energydensity}
    \tilde{\mathcal{E}}=\left|\tilde{\nabla}\Theta\right|^{2}+\sin^{2}\Theta\left|\tilde{\nabla}\Phi\right|^{2}-\cos^{2}\Theta-2h\sin\Theta\cos\Phi.
\end{equation}

Notice that, as a consequence of our local approximation for the magnetostatic energy, the exchange length (Eq.~\ref{eq:exchangelength}) depends on the disk radius. The implicit definition is not a problem for computation because $\zeta(R)$ is monotonic, and therefore, $l_\mathrm{ex}(\zeta(R))$ is single-valued. The shape-induced effects of the local approximation into the effective anisotropy are incorporated in this definition to smooth the transition between the macrospin and the wall-mediated models.

We will assume that $\Phi$ is constant ($\Phi=0$) and the profile $\Theta$ only depends on $x$. 
\begin{equation}
    \tilde{E}(\Theta(x))=\int\left[ \left|\frac{\partial\Theta}{\partial \tilde{x}}\right|^{2}-\cos^{2}\Theta-2h\sin\Theta 
    \right]
    d^{2}\mathbf{\tilde{r}}.
\end{equation}
Here, we maintain a two-dimensional measure of integration to account for the circular shape of our device, even though the $y$ coordinate is ignored when performing variational calculus. This is the key approximation in our model. We solve for the transversal profile of an infinite stripe and assume that it is still valid for a circle.

The configurations $\Theta(x)$ that correspond to energy minima
or saddle points are extremes of this function which can be found
from the corresponding Euler-Lagrange equation and its accompanying boundary condition (see App. \ref{app:derivationofboundaryconditions}).

\begin{equation}
    \frac{\partial^{2}\Theta}{\partial \tilde{x}^{2}}=\cos\Theta\left[\sin\Theta-h\right],
    \qquad
    \left.\frac{\partial \Theta}{\partial \tilde{x}}\right\Vert _{\tilde{x}=\pm \tilde{R}}=0
    .\label{eq:firstvariation}
\end{equation}

A non-uniform solution that satisfies Eq. \ref{eq:firstvariation} exists and corresponds to a reversal driven by a domain wall that moves across the disk (detailed derivation in Appendix~\ref{app:derivationnonconstantsaddle}). The key observation is that the saddle state occurs when the domain wall is at the center of the disk, and the magnetization profile described by:
\begin{equation}
    \Theta_0(\tilde{x})\equiv\frac{\pi}{2}-2\tan^{-1}\left[\sqrt{m}\  \mathcal{B}\ 
    \mathrm{sn}\left(\left.\frac{\tilde{x}}{\mathcal{A}}
    \right|m\right)\right].\label{eq:instantonsaddle}
\end{equation}
In the equation, $\mathrm{sn}(\cdot|m)$ is the Jacobi Elliptic function with parameter $m$ chosen to satisfy the boundary conditions (not to be confused with the magnetization magnitude $|\mathbf{m}|$=1). Appendix~\ref{app:profileparameters} explains how to obtain the parameters used in the description of these profiles, i.e., $\mathcal{A}$, $\mathcal{B}$, and $m$.

Numerically integrating the energy density (Eq.~\ref{eq:energydensity}) over the surface of the disk, 
\begin{equation}
    E= A t\int_{-\tilde{R}}^{\tilde{R}} 2 \sqrt{\tilde{R}^2-\tilde{x}^2}\tilde{\mathcal{E}}(\Theta_0 (x))d\tilde{x},
\end{equation}
gives the energy of the non-uniform saddle state. The energy barrier is therefore:
\begin{equation}
    E_b=E-At\pi\tilde{R}^2\tilde{\mathcal{E}}_1.
\end{equation}

\subsection{\label{sec:stringMeth} String Method Calculations}

The String method \cite{eSimplifiedImprovedString2007,eStringMethodStudy2002} is a numerical procedure for calculating transition energies and paths within the context of large fluctuations and rare events and it is especially useful to find the minimum energy path (MEP) connecting two metastable configurations. Practically, it is a chain-of-states algorithm \cite{bessarab_method_2015} that allows for precise estimates of $E_b$ when analytical solutions of the saddle state cannot be obtained. To use it, we provide a guess for the optimal escape trajectory in configuration space (called a string). Each of the points along this path is a micromagnetic configuration of the disk. If this path is parametrized by the normalized arc-length in magnetization space, the location of each configuration in this path is described by a number from 0 to 1 and is referred to as the reaction coordinate.

\begin{figure}
    \includegraphics[width=\columnwidth]{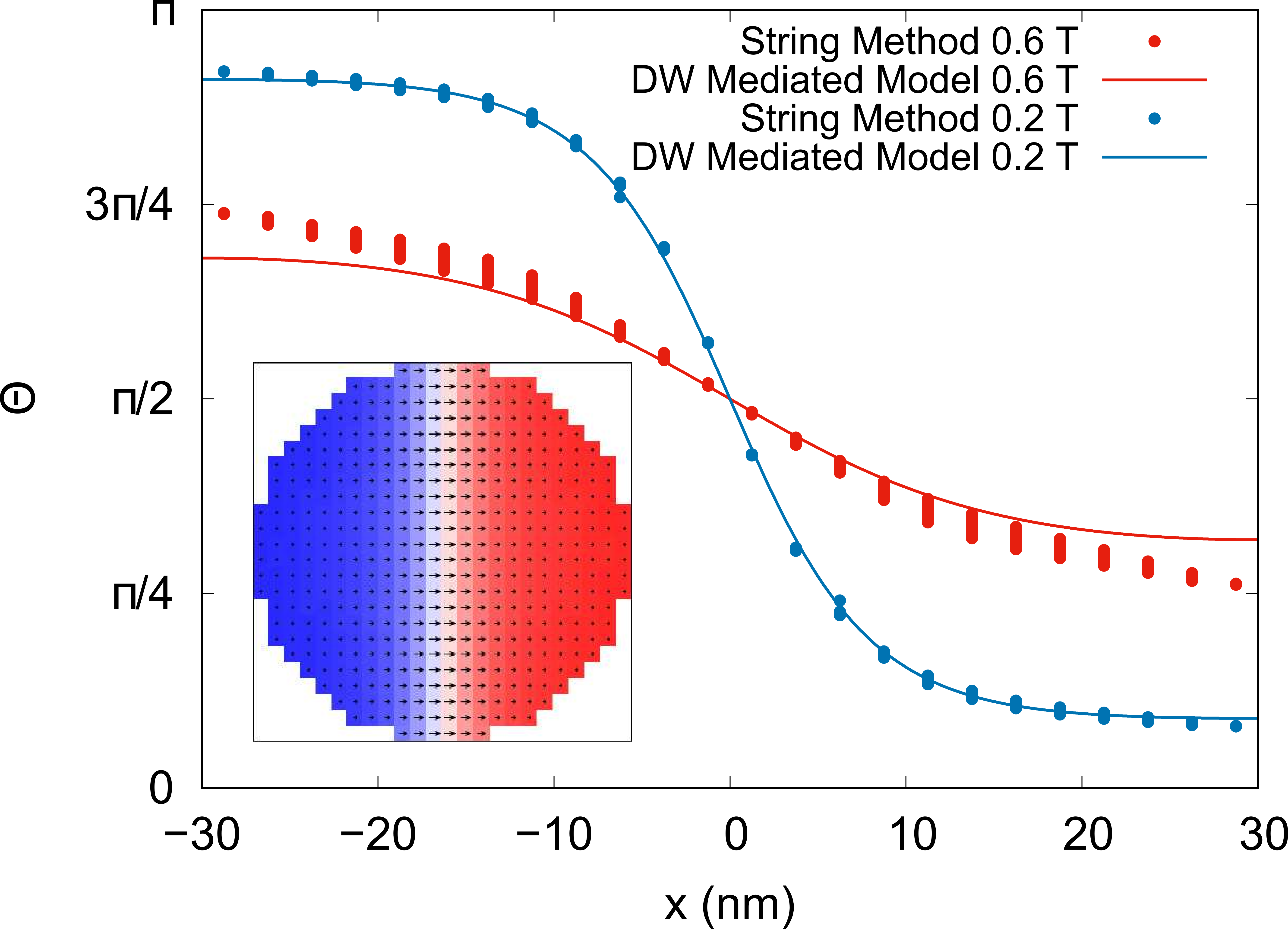}
    \caption{\label{fig:string} Example of energy values along the initial and relaxed string. Configurations along the initial string are shown as snapshots in the upper part of the figure. The string magnetization configurations, after relaxation, are shown at the bottom of the figure. The chosen color palette is linked to the $z$ component of the magnetization (blue for $m_z=-1$; white for $m_z=0$; and red for $m_z=+1$). The saddle configuration is at the center and the two metastable states are far left and far right.}
\end{figure}

We have used our OOMMF-based implementation of this method to find the transition states of the nanodisk at different applied fields. An example of this calculation is shown in Fig.~\ref{fig:string}. The initial trajectory consisted of 100 configurations that contained  domains of zero width ($m_z$ changed abruptly from +1 to -1 between neighboring cells).  The $x$ coordinate of the wall center varies from the left to right of the disk as the reaction coordinate changes from 0 to 1. The top row of thumbnails contains the spin configurations of the initial string. \par
As we can see in Fig.~\ref{fig:string}, the energy of the magnetic configurations along the string starts as a staircase or steplike curve but quickly relaxes to a smooth curve with a single maximum in $\sim 250$ iterations. At the bottom of Fig.~\ref{fig:string}, we present the relaxed spin configuration at different reaction coordinates, which shows the domain wall motion that mediates the switching. The individual configurations are slightly curved Néel walls that sweep the disk from left to right.\par
\begin{figure}
    \includegraphics[width=\columnwidth]{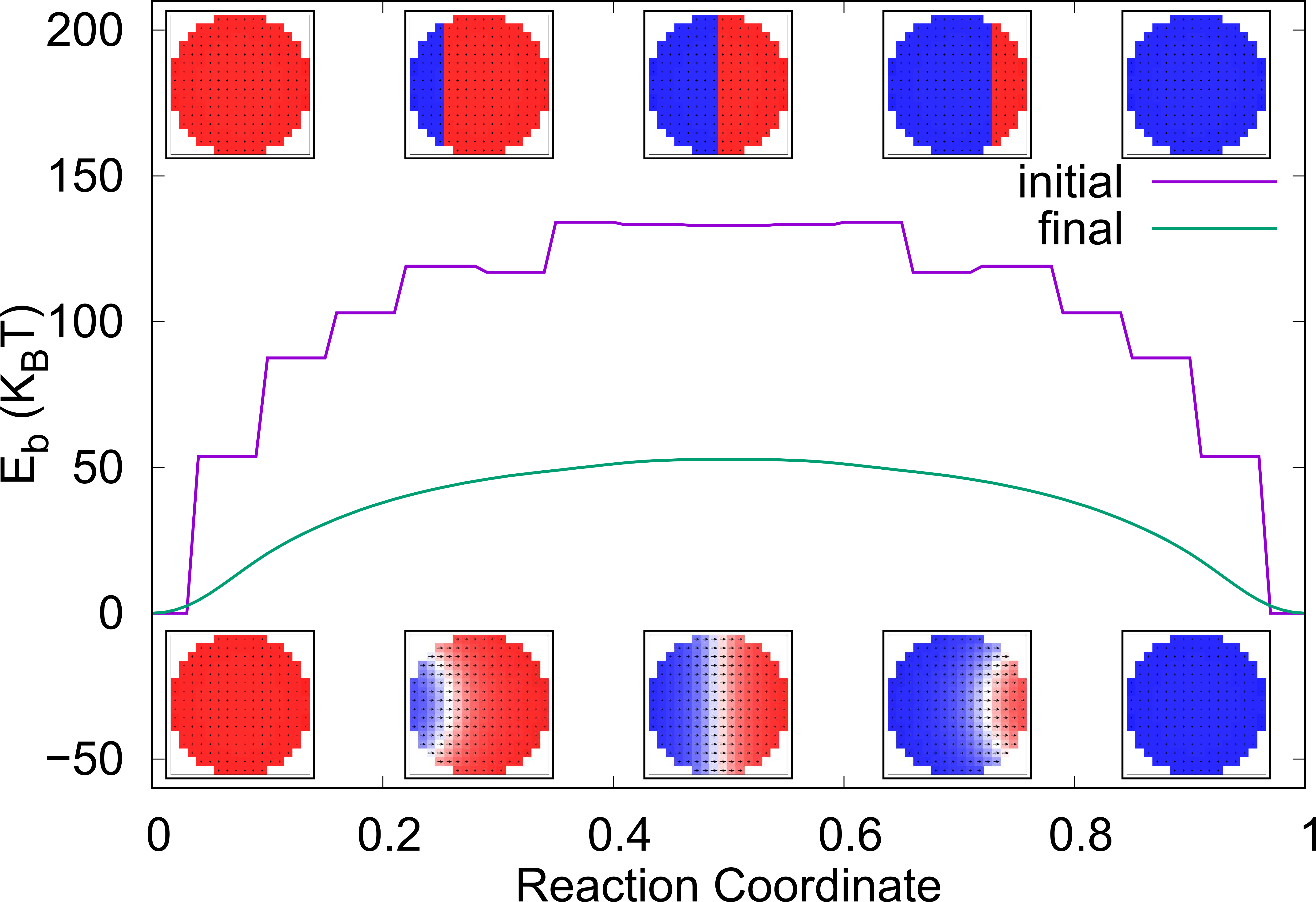}
    \caption{\label{fig:profile} Profile of selected configurations for a \SI{60}{\nano\meter} device. For low fields (\SI{0.2}{\tesla}), the prediction from the domain wall mediated model and the String method simulations match closely. At larger fields (\SI{0.6}{\tesla}), a deviation is visible at the edges of the device, which can be attributed to the fact that our boundary conditions are strictly valid only in infinitely long stripes. Points show the magnetization of each cell obtained from OOMMF. Inset shows the magnetization configuration for the \SI{0.2}{\tesla} case.
    } 
\end{figure}

After the relaxation of the string, the transition between the two domains became smoother with a profile that closely approximates our theoretical predictions (Eq.~\ref{eq:instantonsaddle}), Fig.~\ref{fig:profile}. The String method data is plotted for each cell (symbols) and the scatter measures the deviation of a full micromagnetic calculation from our one-dimensional model. In OOMMF the magnetization profile is allowed to vary in both spatial directions.

 The re-scaled field $\mathbf{h}$ varies between 0 and 1 in our model, and as such determines a critical field, $B_c(R)=\mu_0 M_s(Q-\zeta)$, at which the system becomes saturated in-plane and the bi-stability is lost. As the magnet's aspect ratio changes from a high needle-like cylinder ($\zeta\rightarrow-1/2$) to an extended film $\zeta\rightarrow1$, the field decreases from $B_c=$ \SI{1.7}{\tesla} to \SI{0.61}{\tesla}. Since film devices have small aspect ratios, the fields of interest are all below \SI{0.7}{\tesla}.

\section{\label{sec:results}Results and Discussion}
Since the energy barrier height greatly influences the rate of fluctuation and this barrier can be changed by applying a transverse field to the device, the following section examines the effect of the field on the energy barrier for several disk diameters.

\subsection{\label{sec:field}Size and Field Dependence}
\begin{figure}
    \includegraphics[width=\columnwidth]{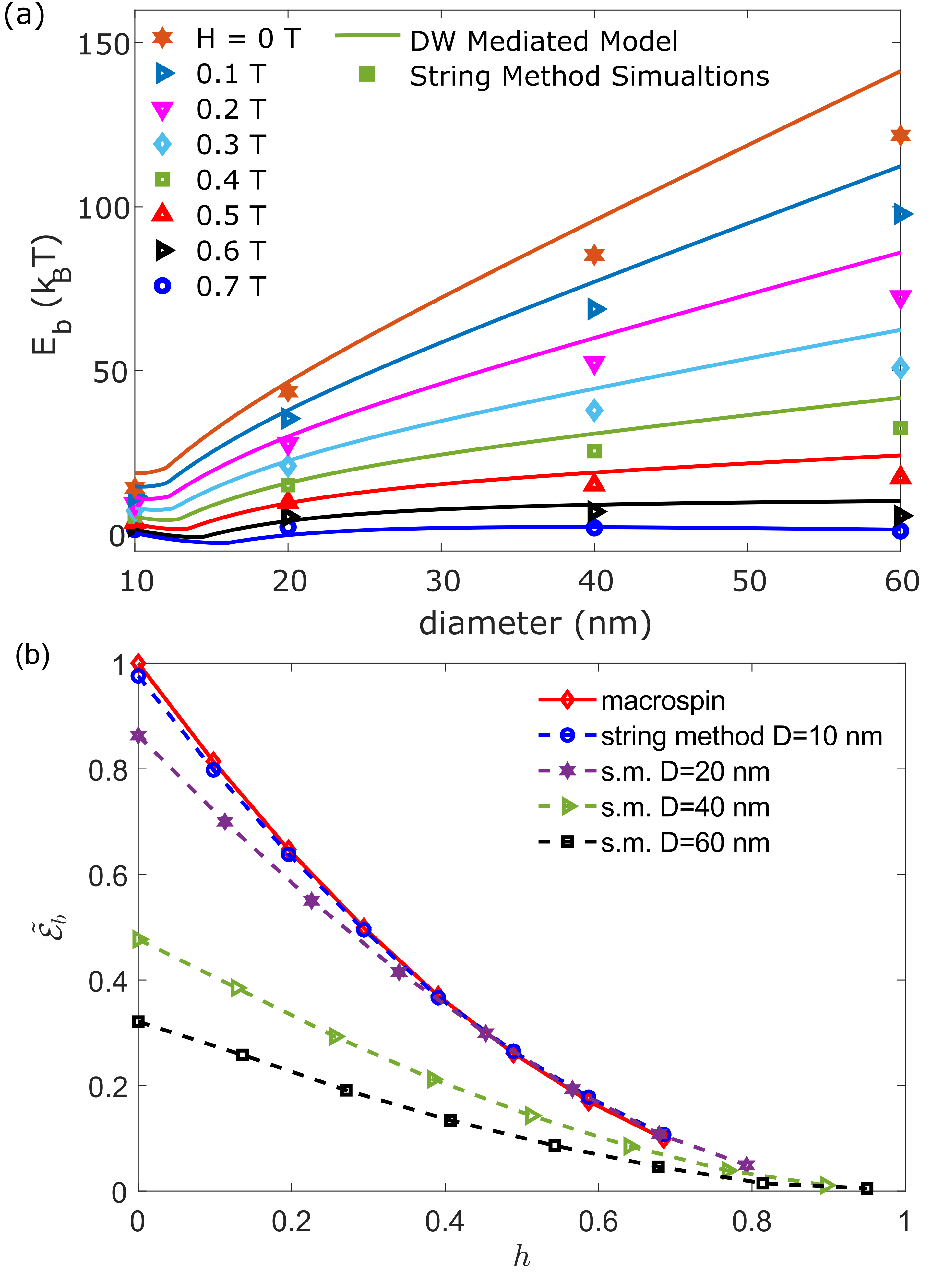}
    \caption{\label{fig:size_dep} (a) Energy barrier ($E_b$) vs size at different fields in a DW-mediated transition. We notice a non-monotonicity vs. field for high fields (i.e. $>$ \SI{0.6}{\tesla}). (b) Scaled energy barrier vs. scaled field. All results obtained from the macrospin model are characterized by the top curve in red.
    }
\end{figure}

In Fig.~\ref{fig:size_dep}(a), we show the energy barrier change with the applied field for different device sizes according to the domain wall (DW) mediated models and the String method calculations. In general, the barrier decreases with increasing field and decreasing diameter. There is, however, a deviation from this trend for high fields as expected from the dependence of the critical field $B_c$ on device radius.

Overall, there is a good agreement between predictions from the DW-mediated model and the String method results although the latter yields slightly lower energy barriers. This is expected since our model only considers a one-dimensional dependence of the magnetization, and the String method calculation allows for other relaxation paths. As a rule, chain-of-states algorithms reach transition paths that are lower in energy than their analytical counterparts since they have larger degrees of freedom and are able to access a wider configurational space. This explains why most of our analytical values are above their corresponding simulation results in Fig.~\ref{fig:size_dep}.

Another result is a non-monotonicity of the energy barrier with respect to nanodisk size, Fig. \ref{fig:size_dep}(a). This is evident for the curves at large fields ($H = 0.6$ and \SI{0.7}{\tesla}) since the macrospin's energy barrier vanishes for $H>\mu_0 M_s(Q-1)$ as $R\rightarrow\infty$.

Figure~\ref{fig:size_dep}(b) shows the scaled energy density barrier vs. the scaled applied field. The dashed lines capture String method calculations at different diameters. The data are the same as those in panel (a) but this new representation helps to easily compare the behavior of nanodisks (of different dimensions) under transverse fields. As disk size decreases towards 10 nm diameter, the energy barrier approaches the value predicted by the macrospin model (Sec.~\ref{sec:macrospinmodel}). 
At \SI{10}{\nano\meter} we have reached the critical dimension for uniform switching, in fact, the curve overlaps perfectly with the macrospin even at zero applied field. For larger disk diameters, a stronger field needs to be employed to have an agreement ($h = 0.4$ for a \SI{20}{\nano\meter} device). The larger disks considered here ($D =$ 40 and \SI{60}{\nano\meter}), do not show macrospin behavior. They exhibit a lower energy barrier for all the applied fields.

\subsection{Tunneling Magnetoresistance Dependence on Field}
\begin{figure}
    \includegraphics[width=\columnwidth]{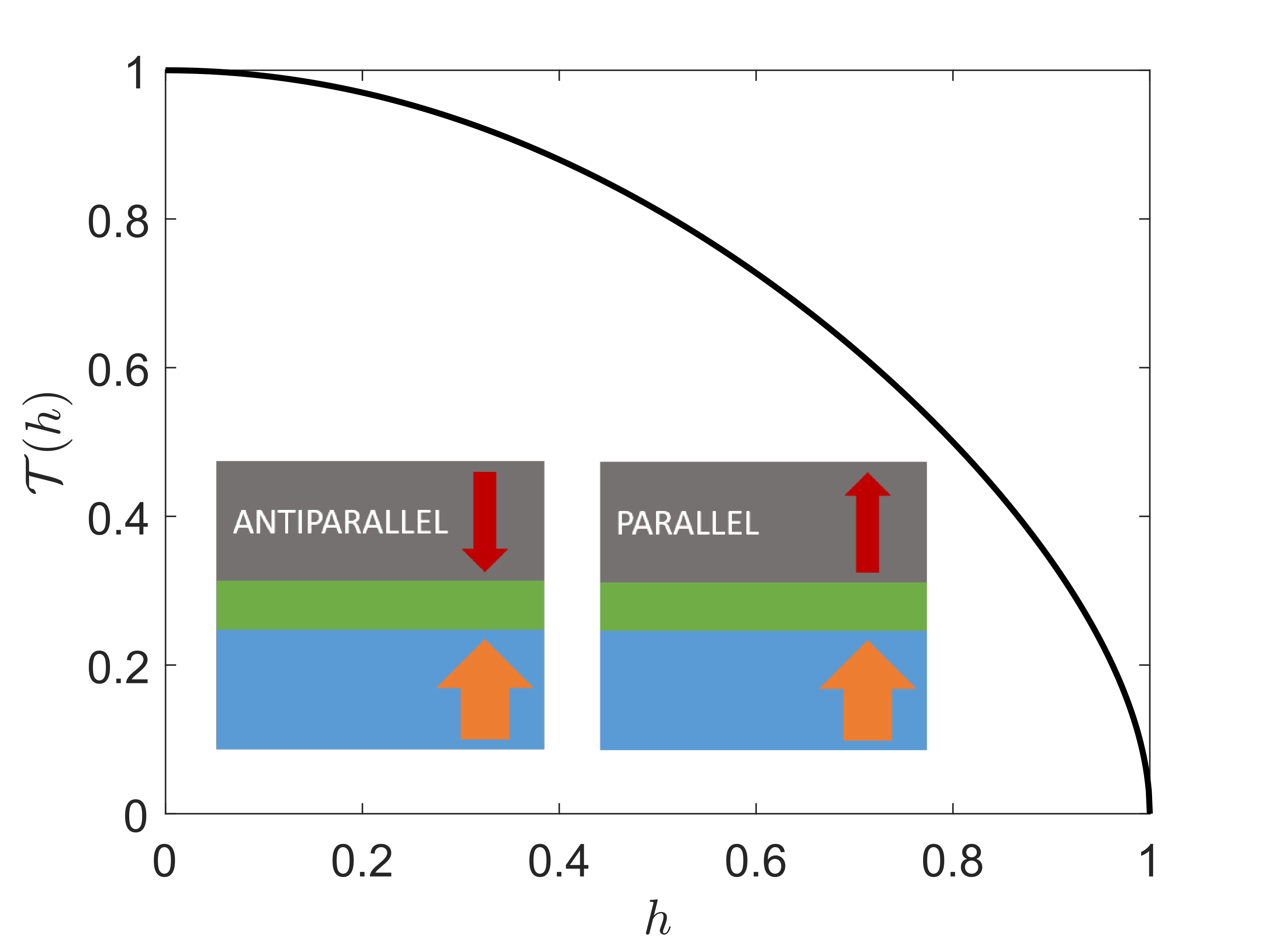}
    \caption{\label{fig:tmr} Tunneling magnetoresistance $\mathcal{T}(h)$ vs field $h$. The magnetoresistance is reasonably large even for large transverse fields, for fields at which the energy barrier can be reduced significantly. In the inset, for the reader's convenience, parallel and antiparallel states schematic in a pMTJ.}
\end{figure}

An expected effect of the application of a transverse field to an MTJ is a decrease in tunneling magnetoresistance (TMR). To predict the TMR vs applied field behavior, we have assumed a fixed reference layer (blue layer in the inset of Fig.~\ref{fig:tmr}).\par
The tunneling conductance is assumed to depend on $\theta$ as follows~(e.g., \cite{zhu_magnetic_2006}):
\begin{equation}
    G(\theta)=\frac{1}{2}(G_P+G_{AP})+\frac{1}{2}(G_P-G_{AP})\cos \theta 
\end{equation}
and the tunneling magnetoresistance is the fraction:
\begin{equation}
    \label{eq:tmr}
    T=\frac{G_P-G_{AP}}{G_{AP}}
\end{equation}
In our case, from Eq.~\ref{eq:theta}, the angle is defined as $\Theta_{\uparrow}=\sin^{-1}h$ and $\Theta_{\downarrow}=\pi-\Theta_{\uparrow}$.

A nice way to see the reduction in TMR is to relate $\mathcal{T}(h)$ to $T $, the TMR as a function of the transverse field to the zero field value (i.e., that given by Eq.~\ref{eq:tmr}). Then the maximum device TMR is given by:
\begin{equation}
    \mathcal{T}(h)=\frac{2T\sqrt{1-h^2}}{T(1-\sqrt{1-h^2})+2}.
\end{equation}
Figure~\ref{fig:tmr} shows a plot of the TMR as a function of the transverse field $h$ with $T = 1.0$, an initial TMR of 100~\%.

\subsection{\label{sec:meta}Intermediate Metastable State}
\begin{figure}
    \includegraphics[width=\columnwidth]{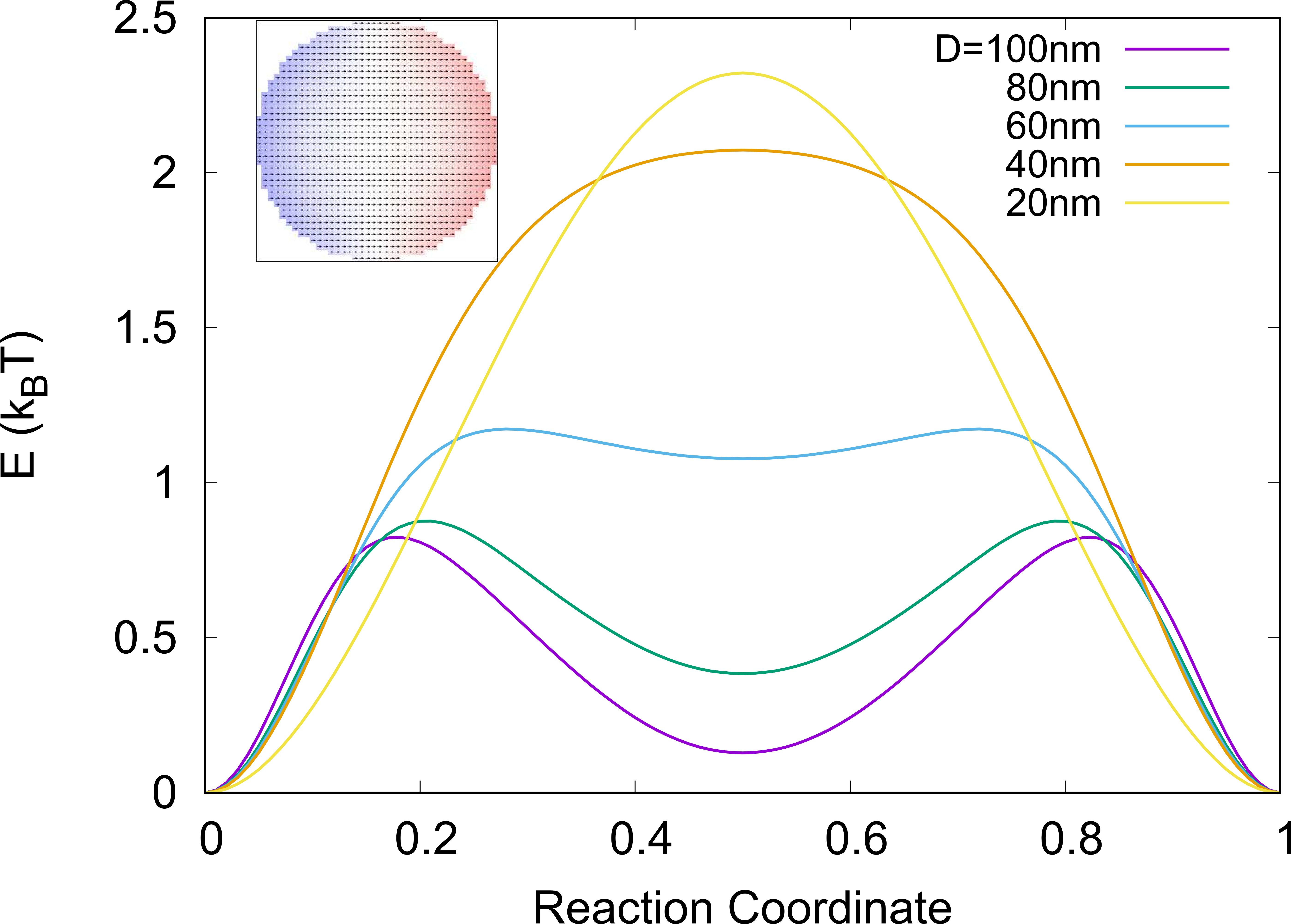}
    \caption{\label{fig:energy} Energy values along relaxed strings for different device sizes with an applied field \SI{0.7}{\tesla}. From $D =$ \SI{60}{\nano\meter} a new metastable state appears at the center of the string and it cannot be predicted by our model. Inset shows the magnetization configuration for $D =$ \SI{100}{\nano\meter} and at a reaction coordinate of 0.5.}
\end{figure}

In our investigation, a new intermediate state appears for large devices ($D \geq$ \SI{60}{\nano\meter}), Fig. \ref{fig:energy}. This new metastable state is higher in energy with respect to the endpoint reaction coordinates. The configuration at this new local energy minimum is highly non-uniform and cannot be predicted from our one-dimensional model that features a narrow domain wall only. We will just provide a heuristic explanation about what may stabilize the metastable state by considering the preferred magnetization direction of an element of volume $td^2r$ depending on its position inside the disk. 

The inset of Fig. \ref{fig:energy} shows the magnetization configuration for $D =$ \SI{100}{\nano\meter} and at reaction coordinate of 0.5. To understand this case, let us first consider the magnetization at the rim of the disk, under the applied field $H\mathbf{\hat{x}}$. At position $x=\pm R$, there is an out-plane buckling of the magnetization as if the effective perpendicular anisotropy became enhanced in the vicinity of the in-plane edge. This out-of-plane orientation decreases the crystalline anisotropy energy at the expense of increasing the Zeeman and magnetostatic energies.

Next, we consider the area closer to the disk center. The penalty for out-of-plane magnetization orientations is stronger in the interior of the disk than at the edges. In the presence of a strong field along $\mathbf{x}$ direction, this magnetization direction becomes favorable. By enlarging the area with close to uniform magnetization, the exchange energy penalty is reduced.

Putting together all these observations, we now provide an explanation for the stability of the newly found configuration. The field is strong, so both metastable directions are slightly out of plane. The perpendicular anisotropy is stronger at the disk edges than in the interior. As a result, the magnetization aligns with the field for most points except within a narrow rim near the disk edge where the contribution to the exchange energy is concentrated.

\section{\label{sec:summary}Conclusions}
In summary, we have compared two possible transition pathways for thermally activated switching, in perpendicularly magnetized nanodisks. One pathway is a coherent rotation of the magnetization (macrospin) and another transition is domain wall-mediated switching with spatially non-uniform magnetization states. These transition paths were tested using overdamped micromagnetic and String method simulations respectively. Specifically, we have compared energy barriers and the spatial profiles of analytical models to those obtained numerically resulting in minimal discrepancies, and only in extreme cases. Moreover, a new metastable state in the energy barrier behavior was found in large devices ($D \geq$ \SI{60}{\nano\meter}) for very high fields.

From an application perspective, the aim of this work was to prove the possibility of using a transverse field to lower the energy barrier of pMTJ devices with different nanodisk diameters. Both the models and the simulations, show that the energy barrier of a perpendicularly magnetized nanodisk can be tuned with the applied field. A quick estimate of the switching rate can be done assuming $\Gamma_0=$ \SI{1}{\giga\hertz} as a characteristic frequency of magnetic excitations. In this case, a barrier $E_b=4.6~\mathrm{k_BT}$ results in switching times in the order of \SI{100}{\nano\second}. For a device with a zero field barrier of $E_b=20~\mathrm{k_BT}$ (i.e. a \SI{12}{\nano\meter} diameter device, with a saturation field of $B_\mathrm{sat}=$ \SI{0.98}{\tesla}) an applied field of \SI{0.51}{\tesla} ($h=0.52$) would produce this fluctuation rate. Similarly, a $40~\mathrm{k_BT}$ device (\SI{18}{\nano\meter} diameter, $B_\mathrm{sat}=$\SI{0.9}{\tesla}) achieves the above rate with a field of \SI{0.60}{\tesla} ($h=0.67$). Thus, as we have illustrated in Fig.~\ref{fig:tmr}, the TMR would still be significant, enabling state determination. The findings confirm that pMTJ devices, whose fluctuation rates are determined by a transverse magnetic field, can produce rapid and measurable fluctuations, presenting exciting possibilities for their application in generating genuine random numbers.

What still remains to be explored are the optimal device dimension, the testing of our theory, and simulations against experimental device measurements. Moreover, an accurate estimation of the prefactor in the attempt rate expression would be needed, to fully assess the stochasticity of the device.

\section*{Acknowledgments}
The research at NYU was supported by the DOE Office of Science (ASCR/BES) Microelectronics Co-Design project COINFLIPS.
This work was partially supported by the Swedish Research Council (VR), project Fundamental Fluctuations in Spintronics, 2017-04196. We also thank the funding agencies Nils and Hans Backmark Foundation (J-2021-2437) and Karl Engvers Foundation for supporting the project. The computations were partially enabled by resources provided by the Swedish National Infrastructure for Computing (SNIC) at High Performance Computing Center North (HPC2N) partially funded by the Swedish Research Council through grant agreement no. 2018-05973.
Gabriel D. Chaves-O'Flynn was supported in part by the National Science Centre Poland under OPUS funding Grant No. 2019/33/B/ST5/02013.

\appendix
\section{Demagnetizing factor for a nanodisk.}\label{app:demagnetizingfactor}
The demagnetization tensor coefficient $N_{zz}$ for a uniformly magnetized thin disk can be calculated \cite{beleggiaEquivalentEllipsoidMagnetized2006} using hypergeometric functions \cite{abramowitzHandbookMathematicalFunctions1965, armitage_elliptic_2006}
\begin{equation}
    N_{zz}=1+\frac{8 R}{3\pi t}-{}_{2}F{_{1}}\left[-\frac{1}{2},\frac{1}{2};2,-\left(\frac{2R}{t}\right)^2\right];
    \label{eq:demagtensor}
\end{equation}
or obtained from any standard micromagnetic solver from the magnetostatic energies of mutually orthogonal saturated states, as was done in \cite{chaves-oflynn_thermal_2015}.
It is easily checked that the values obtained with both approaches are identical.

\section{Profile parameters}\label{app:profileparameters}
The coefficients $\mathcal{A}$ and $\mathcal{B}$ can be obtained from $m$ and $h$ (ignoring nonphysical solutions) as follows:

\begin{align}
\mathcal{A}(m,h)&=\frac
{1-m}
{\sqrt{1+m- \sqrt{h^2 (1-m)^2+4 m}}}
\label{eq:Afromm}\\
\sin(\Theta_R)&=
h+
\frac{m+1}{m-1}
+\sqrt{
h^2
+
\frac{4 m}{(m-1)^2}
}
\label{eq:thetarfromm}\\
\beta&\equiv\frac{1}{2}\left(\frac{\pi}{2}-\Theta\right);
\quad \beta_R=\frac{1}{2}\left(\frac{\pi}{2}-\Theta_R\right)\\
\nonumber\mathcal{B}(h,\Theta_R)&=\tan\left(\beta_R\right)\\
\mathcal{B}(m,h)&=
\sqrt{
\frac{2-h (1-m)+\sqrt{h^2 (m-1)^2+4 m}}
{h (1-m)-2 m-\sqrt{h^2 (m-1)^2+4 m}}
}
\end{align}

The configuration described by Eq. \ref{eq:instantonsaddle} has a wavelength equal to $\lambda=4\mathbf{K}(m)$; to satisfy the boundary conditions, the disk diameter has to be a multiple of half this length $D=n\lambda/2$. We then have an additional relation for the non-uniform saddle $\tilde{R}=n\mathcal{A}(m,h)\mathbf{K}(m)$ for which $m$ can be found numerically. Although large disks will be able to sustain a hierarchy of states with multiple domain walls, we will restrict our analysis to the simplest case $(n=1)$. 
As part of the supplementary materials, we provide a Mathematica notebook to obtain $m, \mathcal{A}$ and $\mathcal{B}$  for specific geometric and material parameters as well as a description of the necessary algebraic steps to obtain the above explicit dependence of $\mathcal{A}$ and $\mathcal{B}$ on $m$ and $h$ [supplementary material - S1].

\section{Derivation of the Boundary Conditions}\label{app:derivationofboundaryconditions}
This is obtained from variational calculus:
\begin{equation}
    \delta \tilde{E}=\iint \delta \left[ \left|\frac{\partial\Theta}{\partial \tilde{x}}\right|^{2}-\cos^{2}\Theta-2h\sin\Theta\right]d\tilde{x}d\tilde{y}.
\end{equation}
To simplify our work, in the text we solve the equation at $y=0$ and assume the same profile is valid for the rest of the disk. In reality, this is the solution for a square of side $2R$. We assume the discrepancy with the solution for a disk will be small
\begin{equation}
2\iint 
\left[
\frac{\partial \Theta}{\partial \tilde{x}}
\frac{\partial\delta \Theta}{\partial \tilde{x}}
+\left[  \sin\Theta\cos\Theta-h\cos\Theta
\right]\delta\Theta
\right]
d\tilde{x}d\tilde{y}.
\end{equation}
Using integration by parts in the first term we obtain:
\begin{equation}
2\iint
\left[
-\frac{\partial^2 \Theta}{\partial \tilde{x}^2}
+\sin\Theta\cos\Theta-h\cos\Theta
\right]\delta\Theta
d\tilde{x}d\tilde{y}.
\end{equation}
\begin{equation}
+2\int^{+\tilde{R}}_{-\tilde{R}}\left[\frac{\partial \Theta}{\partial \tilde{x}}\right]^{\tilde{R}}_{-\tilde{R}}\delta\Theta d\tilde{y} 
\end{equation}
An extremal satisfies $\delta E=0$ for arbitrary $\delta \Theta$; therefore, the first of these integrals provides the differential equation to be solved; the second integral, the boundary conditions to be satisfied $\left.\frac{\partial \Theta}{\partial \tilde{x}}\right\Vert_{\tilde{x}=\pm \tilde{R}}=0$.

\section{Derivation of the non-constant saddle profile.}\label{app:derivationnonconstantsaddle}

We postulate that, in addition to the Neumann boundary conditions in Eq.\ref{eq:firstvariation}, the profiles should also satisfy the following symmetry:
\begin{equation}
    \Theta(\tilde{x}=-\tilde{R})=\Theta_{R}\le\frac{\pi}{2},
    \quad
    \Theta(\tilde{x}=+\tilde{R})=\pi-\Theta_{R}.
\end{equation}
Which is motivated on physical grounds. It is reasonable to expect that the non-uniform extremal is made of two energetically equivalent orientations separated by a transition region in between. We expect that for large disks the magnetization fully rotates between $\Theta_{\uparrow}$ and $\Theta_{\downarrow}$, i.e. $\lim_{R\rightarrow \infty}\Theta_R=\Theta_{\uparrow}$.

Multiplying the differential equation in \ref{eq:firstvariation} by $d\Theta/d\tilde{x}$ and integrating in $x$ results in the equivalent equation:
\begin{equation}
    \frac{1}{2}\left(\frac{d\Theta}{d\tilde{x}}\right)^{2}=\frac{1}{2}\sin^{2}\Theta-h\sin\Theta+C_{1}.\label{eq:firstintegration}
\end{equation}
We solve for the first constant of integration, $C_1$, after substitution of the boundary conditions.
\begin{equation}
    C_{1}=\sin\Theta_{R}\left[h-\frac{1}{2}\sin\Theta_{R}\right]\qquad\lim_{R\rightarrow\infty}C_1=\frac{h^2}{2}.
\end{equation}
We use $m_x=\sin(\Theta)$ to rewrite Eq.\ref{eq:firstintegration} as an elliptical integral:
\begin{equation}
    d\tilde{x}=\frac{dm_{x}}{\sqrt{1-m_{x}^{2}}\sqrt{m_{x}^{2}-2hm_{x}+2C_{1}}}.
\end{equation}
After integrating we obtain an expression involving, $F(\tilde{x}|m)$, the Elliptic Integral of the first kind:

\begin{align}
    \tilde{x}+C_2 & =\mathcal{A}\cdot F\left(\left.\sin^{-1}\left(\mathcal{B}\cot\left(\beta\right)\right)\right)|m\right)\label{eq:secondintegration}
\end{align}
Solving Eq.(\ref{eq:secondintegration}) for $\Theta$, we get an expression involving a Jacobi elliptic function, $\mathrm{sn}(x|m)$:
\begin{equation}
    \Theta=\frac{\pi}{2}-2\cot^{-1}\left[\frac{1}{\mathcal{B}}\mathrm{sn}\left(\left.\frac{\tilde{x}+C_2}{\mathcal{A}}\right|m\right)\right].
\end{equation}
Since we expect $\Theta(x=0)=\pi/2$, we can find $C_2$ that satisfies:
\begin{equation}
    \mathrm{sn}\left(\left.\frac{C_2}{\mathcal{A}}\right|m\right)=\infty.
\end{equation}
Since the first pole of $\mathrm{sn}(x|m)$ is located at $i\mathbf{K}(1-m)$, we find $C_2=i\mathcal{A}\mathbf{K}(1-m)$. This allows us to use a change of variables and other   identities of Jacobi elliptic functions and simplify this to Eq. \ref{eq:instantonsaddle}

In the large disk limit ($m$=1), the above parameters can be simplified:
\begin{align}
    \lim_{R\rightarrow \infty} \mathcal{A} & = 
    \sqrt{
    \frac{4}{1-h^2}
    }\\
    \lim_{R\rightarrow \infty}\mathcal{B}&=\sqrt{\frac{1-h}{1+h}}=\frac{1}{\cot{\beta_{\mathrm{max}}}}\\
    \beta_{\mathrm{max}}&\equiv\frac{1}{2}\left(\frac{\pi}{2}-\Theta_{\uparrow}\right)
\end{align}
and, when $R\rightarrow \infty$,
\begin{equation} \Theta_0=\frac{\pi}{2}-2\tan^{-1}\left[\tan\beta_{\mathrm{max}}
    \tanh\left(x \sqrt{\frac{1-h^2}{4}}
    \right)\right].
\end{equation}
In this form, the solution provides for a quick intuitive interpretation of this result. Since $\tanh(x)$ changes sigmoidally from -1 to 1 as x grows from $-\infty$ to $+\infty$, the angle $\Theta$ acquires the same sigmoidal dependence but varies between $\Theta_{\uparrow}$ and $\Theta_{\downarrow}$.     






\bibliography{apssamp.bib}

\end{document}